\documentclass[]{raa}            

\usepackage{graphicx,times}             
\usepackage{csvsimple}
\usepackage{amssymb }
\usepackage{natbib  }

\begin{document}

   \title{Properties of filament in Solar cycle 20-23 from McIntosh database}

 \volnopage{ {\bf 20XX} Vol.\ {\bf X} No. {\bf XX}, 000--000}
   \setcounter{page}{1}

   \author{Rakesh Mazumder \inst{1}
     }

   \institute{$^{1}$Center of Excellence in Space Sciences India, Indian Institute of Science Education and
Research Kolkata, Mohanpur 741246, West Bengal, India \\ 
\vspace{5mm}
   {\small Received 20XX Month Day; accepted 20XX Month Day}
}

\abstract{ Filament is a cool, dense structure suspended in the solar corona. The eruption of a filament is often associated with coronal mass ejection (CME), which has an adverse effect on space weather. Hence, the study of filament has attracted much attention in the recent past. The tilt angle of active region (AR) magnetic bipoles is a crucial parameter in the context of the solar dynamo. It governs the conversion efficiency of the toroidal magnetic field to poloidal magnetic field. The filament always forms over the Polarity Inversion Lines (PILs). So the study of tilt angles of the filament can provide valuable information about generation of magnetic field in the Sun. We study the tilt angle of filaments and other properties of it using McIntosh archive data. We fit a straight line to each filament to estimate its tilt angle. We study the variation of mean tilt angle with time. The latitude distribution of positive tilt angle filaments and negative tilt angle filaments reveal that there is a dominance of positive tilt angle filaments in the southern hemisphere and negative tilt angle filaments dominate in the northern hemisphere. We study the variation of the mean tilt angle for low and high latitude separately. Study of temporal variation of filament number reveals that total filament number and low latitude filament number varies cyclically, in phase with the solar cycle. The number of filaments in high latitude is less, and they also show a cyclic pattern in temporal variation. We also study the north-south asymmetry of filament for different latitude criteria.
\keywords{Sun: filaments, prominences, Sun: magnetic fields, Sun: corona, Sun: activity, (Sun:) sunspots
}
}

   \authorrunning{Mazumder}            
   \titlerunning{Properties of filament from McIntosh database}  
   \maketitle

%
\section{Introduction}           
Filaments are cool and dense structures suspended in the solar corona.  They appear as dark filamentary structures when seen on disk of the Sun. The same feature appears bright when observed in the off limb region and are known as prominence for historical reasons. Filament always appears in the vicinity of Polarity Inversion Lines (PILs) (also known as neutral lines) \citep{1998SoPh..182..107M}. Filament eruption is often associated with coronal mass ejection (CME), which is hazardous for space weather. Hence, the study of the filament is important from the perspective of space weather research \citep{2003ApJ...586..562G,2000ApJ...537..503G,2004ApJ...614.1054J}. Filaments are primarily observed in H$\alpha$ line. Systematic H$\alpha$ observations have started since 1915 at the Kodaikanal Observatory (India), 1919 at the Meudon Observatory (France) (spectroscopic observations), and 1959 at the Kislovodsk Mountain Astronomical Station of the Main (Pulkovo) Astronomical Observatory of Russian Academy of Sciences. \cite{2016SoPh..291.1115T} has studied more than eight solar cycles' data of filaments using Meudon observatory and Kislovodsk mountain station synoptic maps. \cite{2017ApJ...849...44C} has studied filament properties using Kodaikanal Observatory (India) data for nine solar cycles. \cite{2015ApJS..221...33H} has studied properties of filaments using data from Big Bear Solar Observatory (BBSO) from 1988 to 2013 and \cite{2018ApJ...868...52M} has studied properties of filaments using McIntosh archive data from 1967 to 2009.

The filaments are distributed uniformly in longitude \citep{2018ApJ...868...52M}. However, the distribution of filaments in latitude is not uniform. A bimodal nature in latitudinal distribution of filaments is observed \citep{2015ApJS..221...33H,2018ApJ...868...52M}. \cite{2016SoPh..291.1115T}, \cite{2017ApJ...849...44C}, \cite{2015ApJS..221...33H}, and \cite{2018ApJ...868...52M} have reported a butterfly structure in the  temporal variation of the latitudinal distribution of filaments. The butterfly diagram is similar to that of sunspot except that the spread in latitude is broader in case of filaments. They have also reported the rush to the pole of filaments during the maxima of the solar cycle. The variation of filament number is found to be cyclic in close proximity to sunspot number variation \citep{2015ApJS..221...33H,2010NewA...15..346L}. The total filament length in the Carrington map is also found to vary cyclically with time \citep{2016SoPh..291.1115T,2018ApJ...868...52M}. \cite{2018ApJ...868...52M} further classified the filaments as associated and unassociated to solar activity belt according to their position inside or outside the sunspot activity belt respectively. They have found the variation in total length of the filaments is correlated with variation in sunspot area. However, the temporal variation of length of filaments unassociated with solar activity belt is uncorrelated with variation of sunspot area. \\

The tilt angle of the active region (AR) is a very crucial parameter in the solar dynamo. The tilt angle of active region (AR) magnetic bipoles governs the conversion efficiency of the toroidal magnetic field to poloidal magnetic field \citep{2013MNRAS.432.2975T}. Filaments always form over PILs. So the analysis of the tilt angle of filaments can give us important insights into the mechanism of the generation of the magnetic field of the Sun. Average tilt angle of filaments are found to decrease from the equator towards the pole for both positive tilt and negative tilt \citep{2016SoPh..291.1115T,2018ApJ...868...52M} which is consistent with famous Joy's law of sunspot pair \citep{1919ApJ....49..153H}. The tilt angle of filament was observed to be negative in the northern hemisphere and positive in the southern hemisphere for different length and sunspot belt association criteria by \cite{2018ApJ...868...52M}. Following this observation, they concluded that the origin of most of the filaments are from large-scale magnetic field structure and not from the active region (AR). However, the geometry of the inter-AR filament as depicted by \cite{2001ApJ...558..872M} can also contribute to this apparently unexpected tilt angle orientation in both hemispheres. The hemispheric asymmetry of the filament number was studied, and they were found to be asymmetric at different phases of the solar cycle \citep{2018ApJ...868...52M,2015ApJS..221...33H,2015RAA....15...77K,2010NewA...15..346L,2003NewA....8..655L}.

The mean tilt angle variation with time is observed to follow a solar cycle like variation \citep{2016SoPh..291.1115T}. The latitudinal distribution of positive and negative tilt angles of filaments are found to behave differently \citep{2016SoPh..291.1115T}. The time latitude distribution of the filaments' tilt angle shows a positive tilt dominance in mid-latitude and a negative tilt dominance in higher latitude \citep{2016SoPh..291.1115T}. \\

We have organized the paper as follows. In section 2, we describe the data used and discuss the methods implemented to analyze the data. In section 3, we analyze the results obtained. In section 4, we summarize our work.

\begin{figure}[htb!]
\centering
\includegraphics[scale=0.08,angle=0]{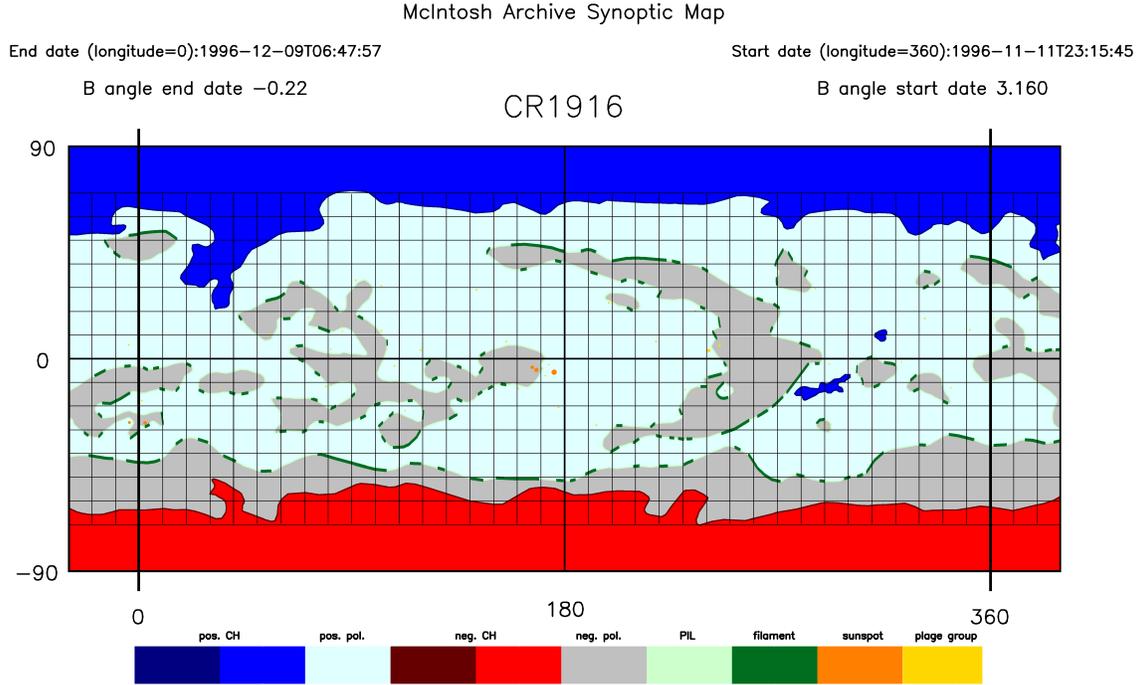}
\caption{Level 3 image of Carrington rotation number 1916 in McIntosh archive. Here, gray and light blue patches represent the negative and positive polarity magnetic fields respectively. The red and dark blue patches represent the negative and positive coronal holes respectively. Green and light green lines represent the filaments, and Polarity Inversion Lines (PILs) respectively.}
\label{level3}
\end{figure}

\begin{figure}[htb!]
\centering
\includegraphics[scale=1.,angle=90]{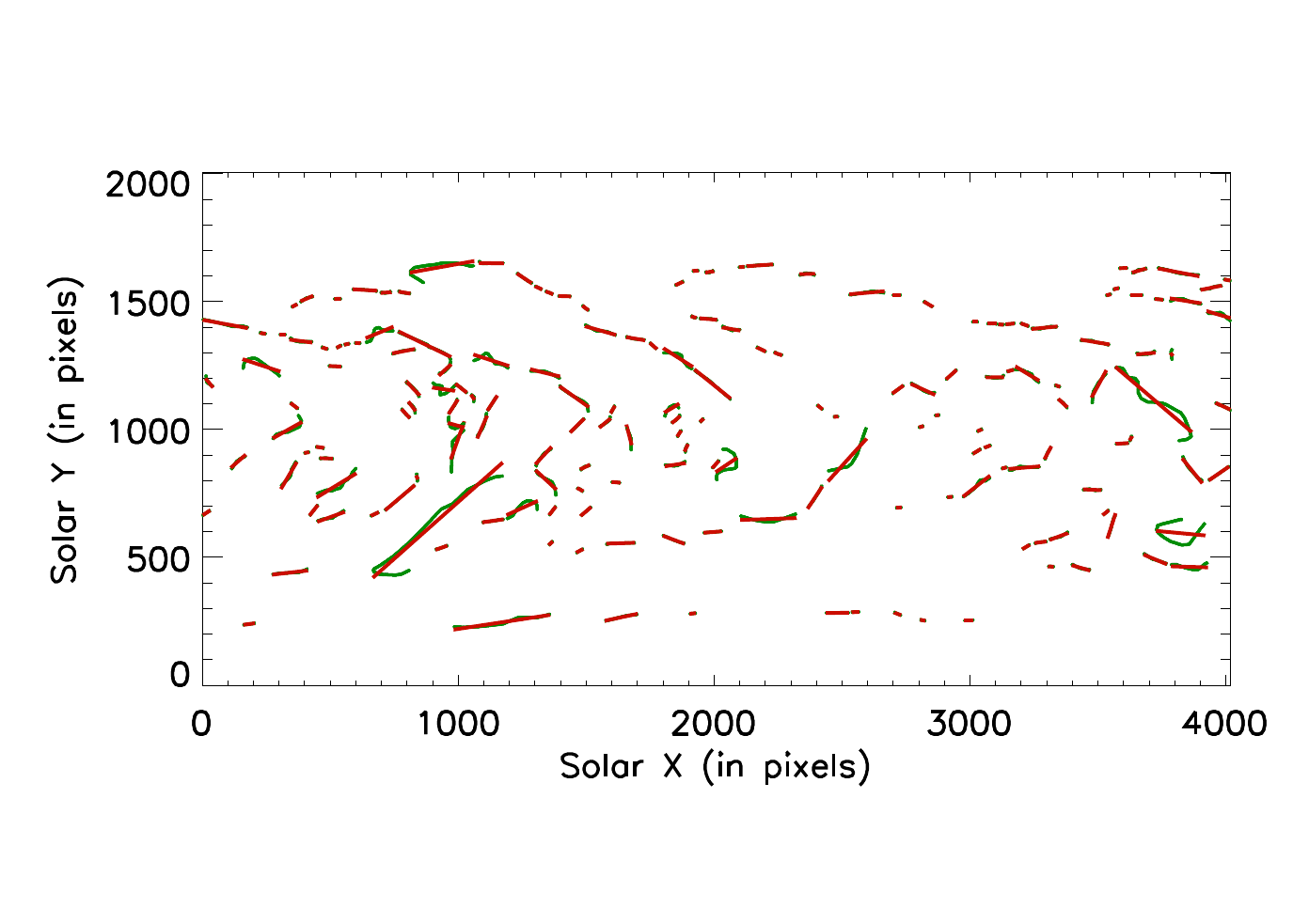}
\caption{Fitting of filaments by straight lines in a sample Carrington map (Carrington number 1552). The green lines represent filaments. The red lines represent fitted straight lines. }
\label{sample_fil_fit}
\end{figure}

\begin{figure}[htb!]
\centering
\includegraphics[scale=1.1,angle=90]{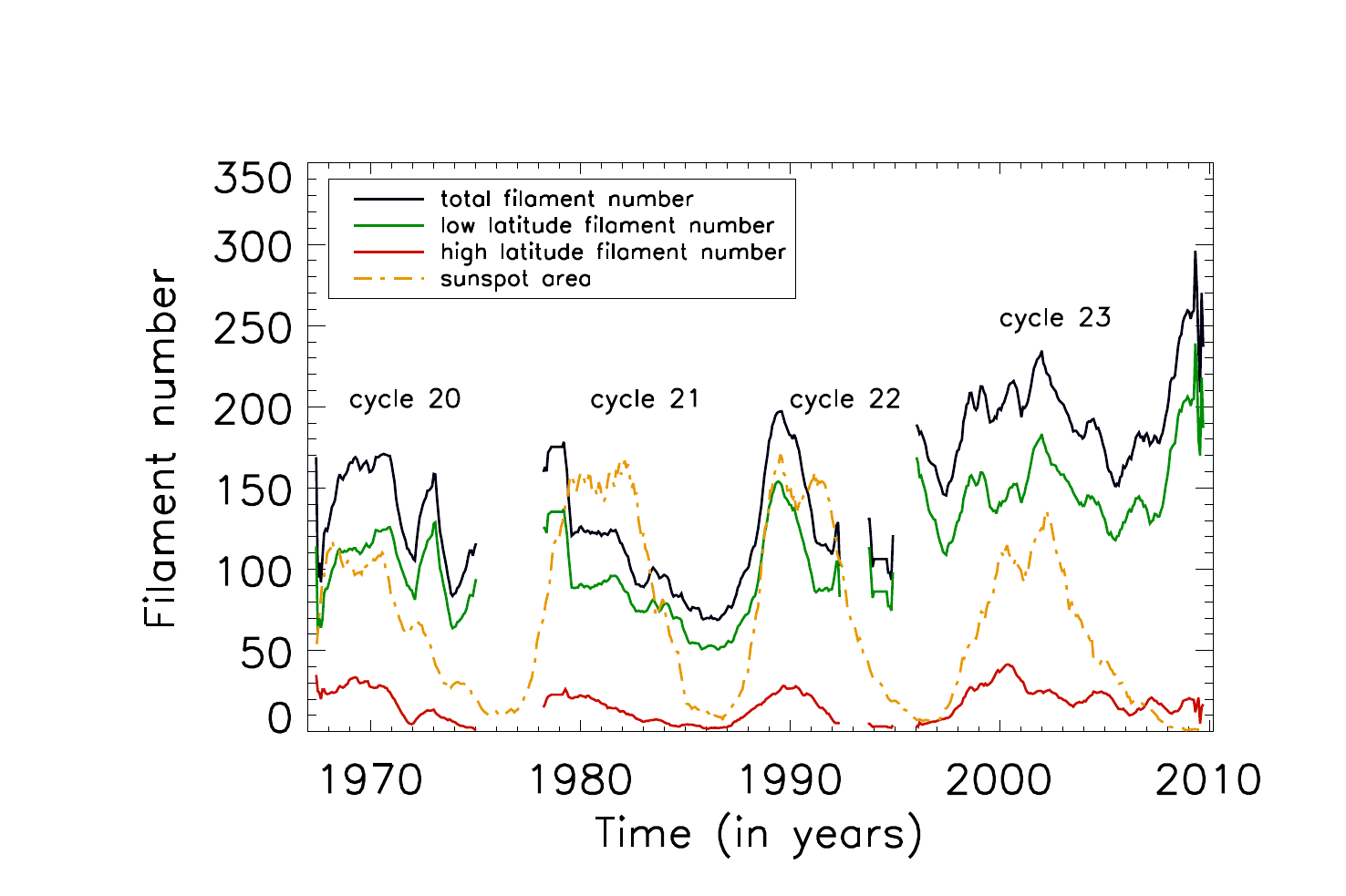}
\caption{Filament number variation with time. The black line shows the variation in total number of filament with time. The green line depicts the variation of low latitude filament number with time. The red line represents the variation of high latitude filament number with time. The sunspot area variation is depicted by the orange dashed line.}
\label{fil_no}
\end{figure}

\begin{figure}[htb!]
\centering
\includegraphics[scale=0.3,angle=0]{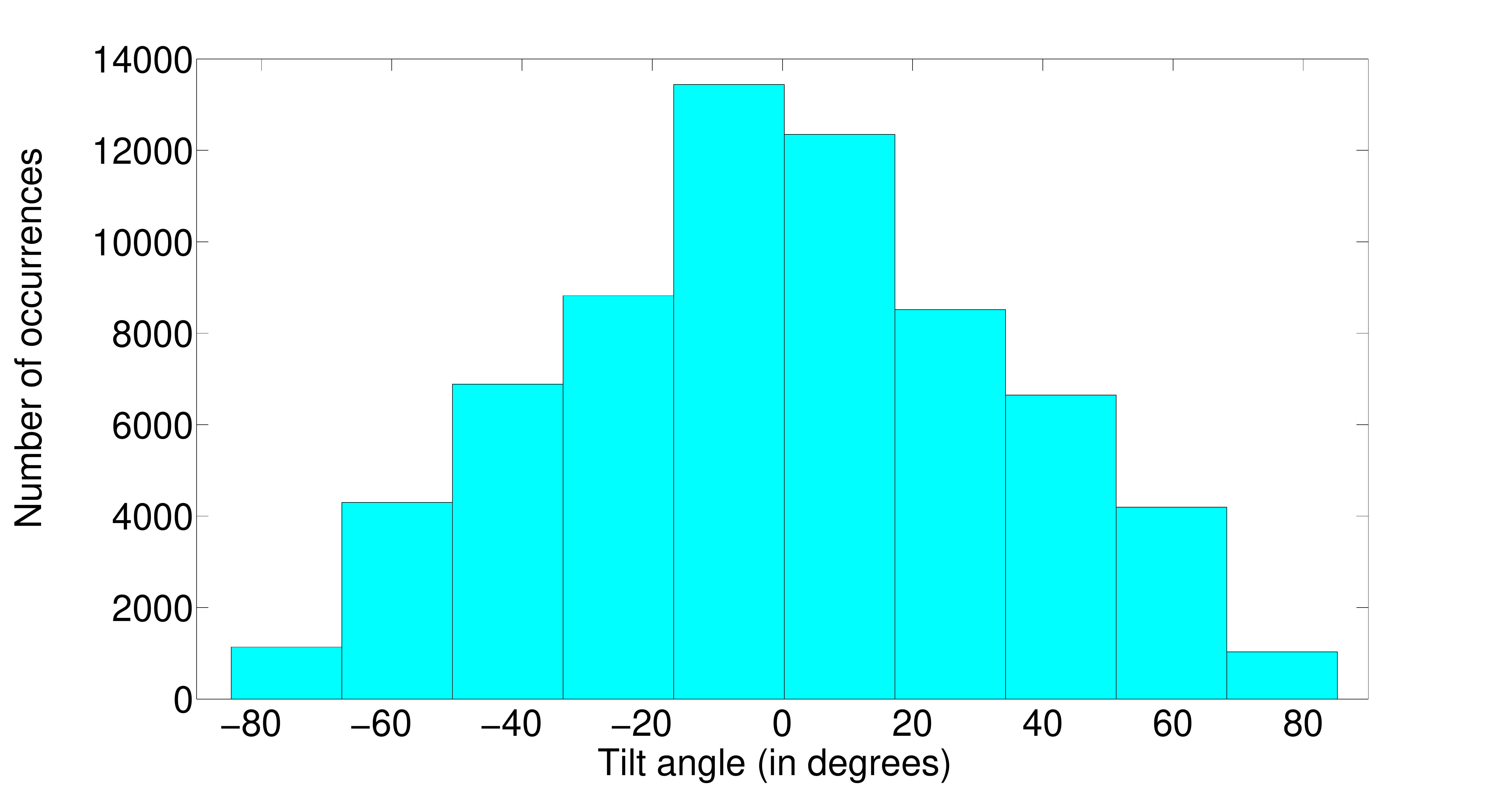}
\caption{Histogram of the tilt angle of filaments. The negative tilt angle is excess in comparison to positive tilt angle. }
\label{hist_tilt}
\end{figure}

\begin{figure}[htb!]
\centering
\includegraphics[scale=0.52,angle=90]{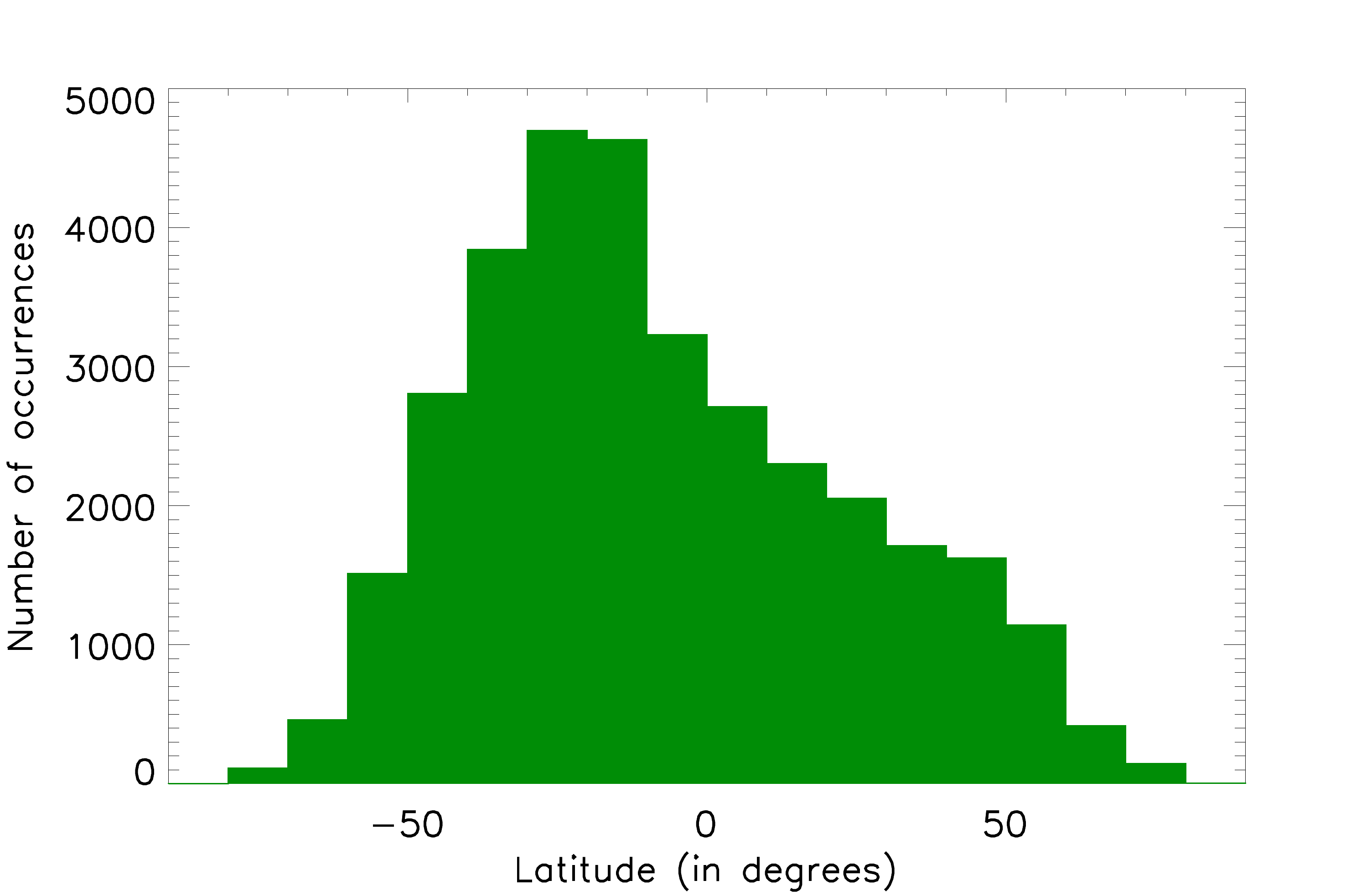}
\caption{Histogram of latitude of filaments having positive tilt angle. Southern hemisphere has more positive tilt angle filaments than northern hemisphere.}
\label{hist_pv}
\end{figure}

\begin{figure}[htb!]
\centering
\includegraphics[scale=0.52,angle=90]{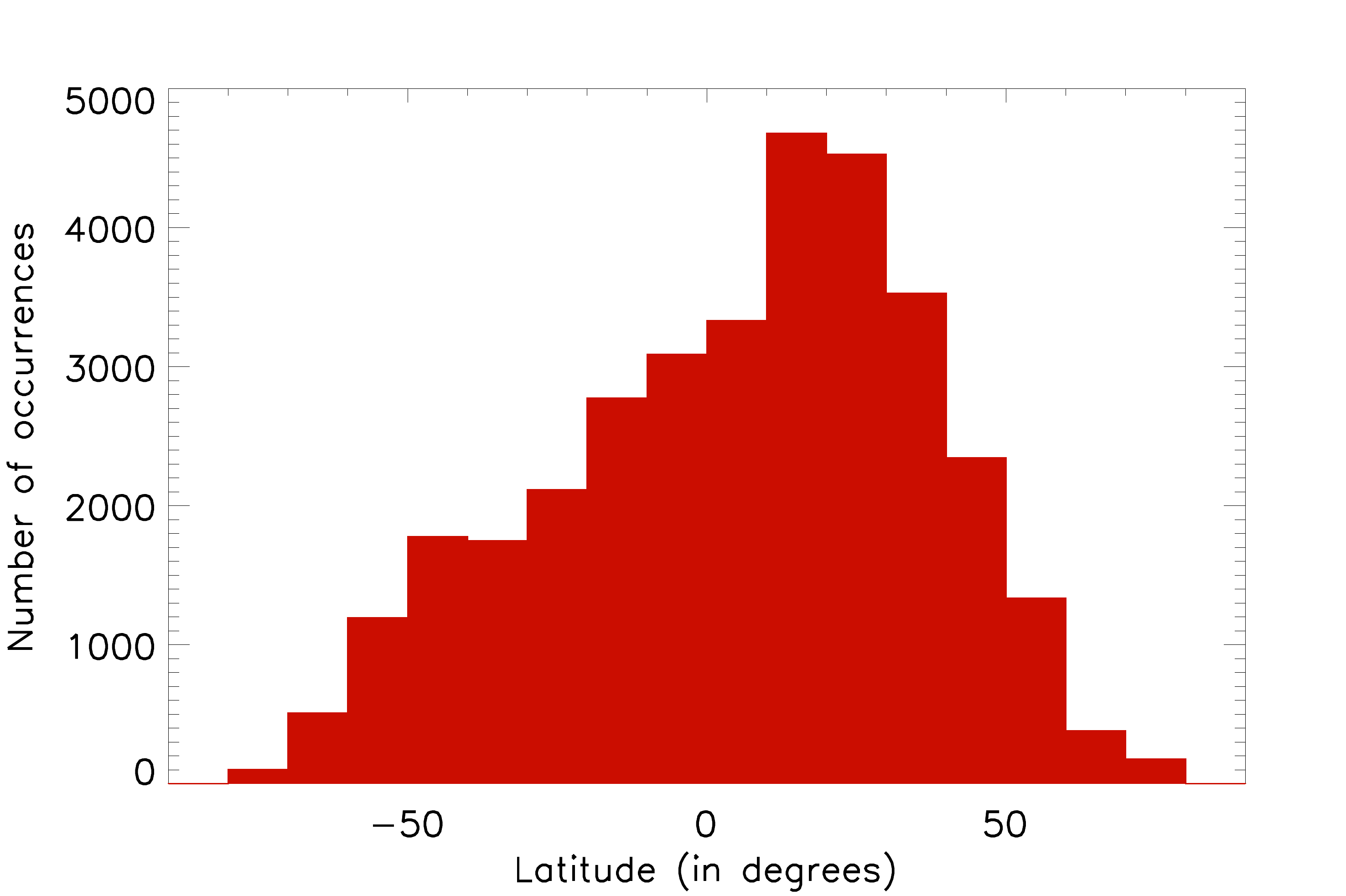}
\caption{Histogram of the latitude of filaments having negative tilt angle. Northern hemisphere has more negative tilt angle filaments than southern hemisphere.}
\label{hist_nv}
\end{figure}

\begin{figure}[htb!]
\centering
\includegraphics[scale=0.5,angle=90]{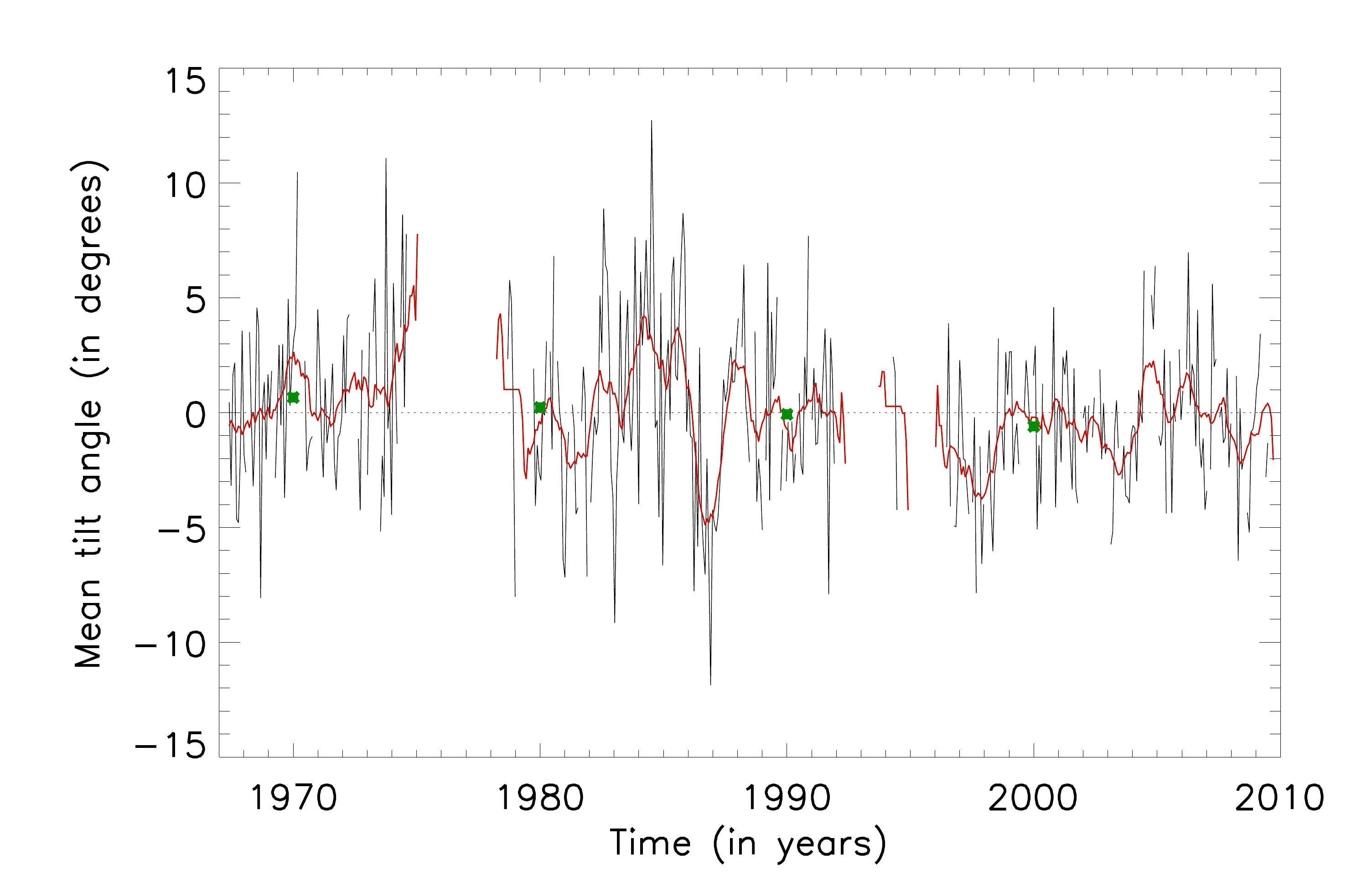}
\caption{ Mean tilt angle variation with time. The black line shows the time variation of the  mean tilt angle.  The red line depicts temporal variation of the mean tilt angle averaged over 13 Carrington rotations. The green cross represent solar cycle average mean tilt angle.}
\label{tilt_avg}
\end{figure}
 
\begin{figure}[htb!]
\centering
\includegraphics[scale=0.5,angle=90]{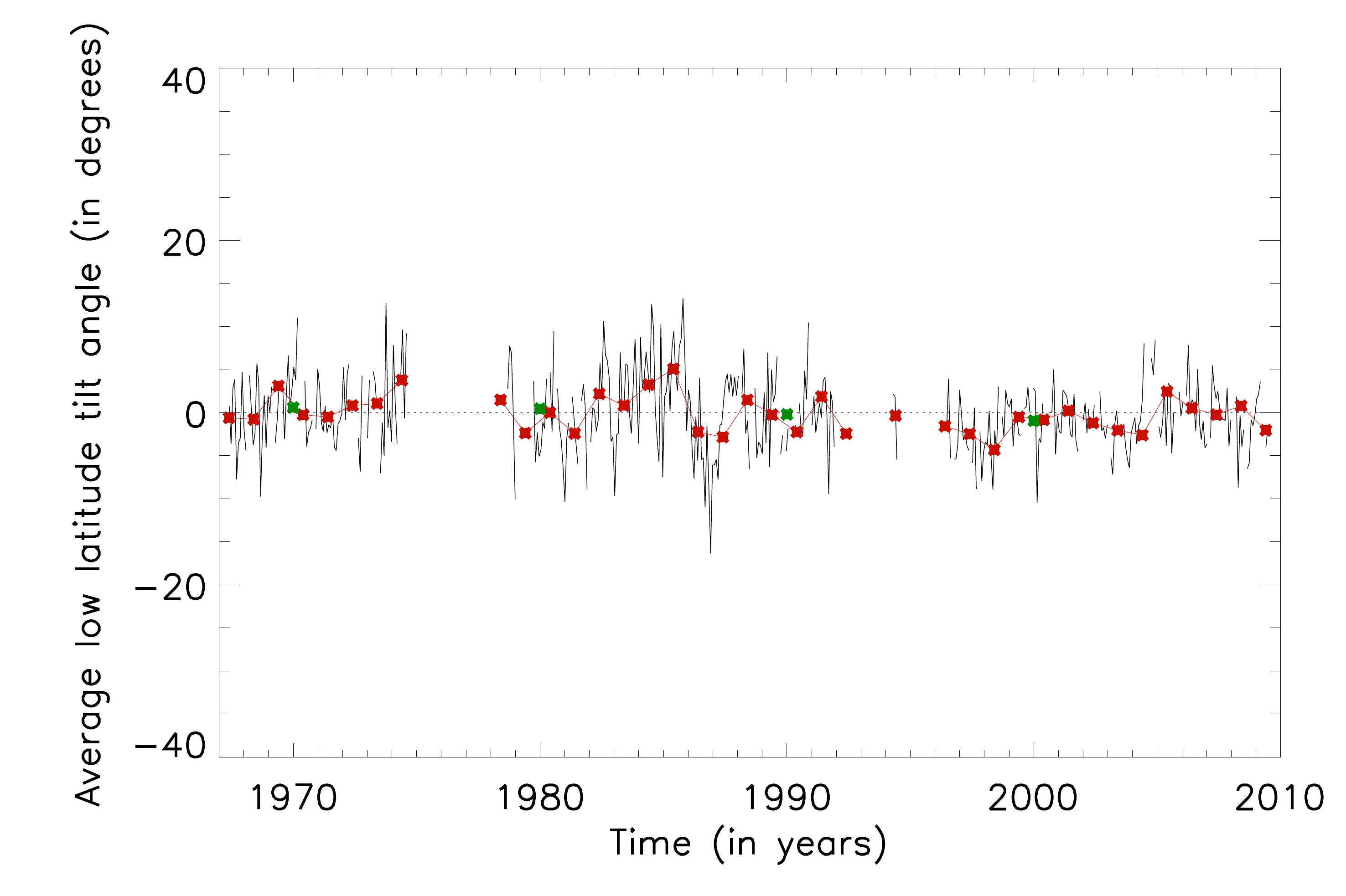}
\caption{Mean tilt angle variation of low latitude filament with time. Black line depicts the low latitude mean tilt angle variation with time. The red line shows the temporal variation of low latitude mean tilt angle averaged over 13 Carrington rotations. The green cross represents solar cycle average mean tilt angle.}
\label{tilt_avg_lw}
\end{figure}

\begin{figure}[htb!]
\centering
\includegraphics[scale=0.5,angle=90]{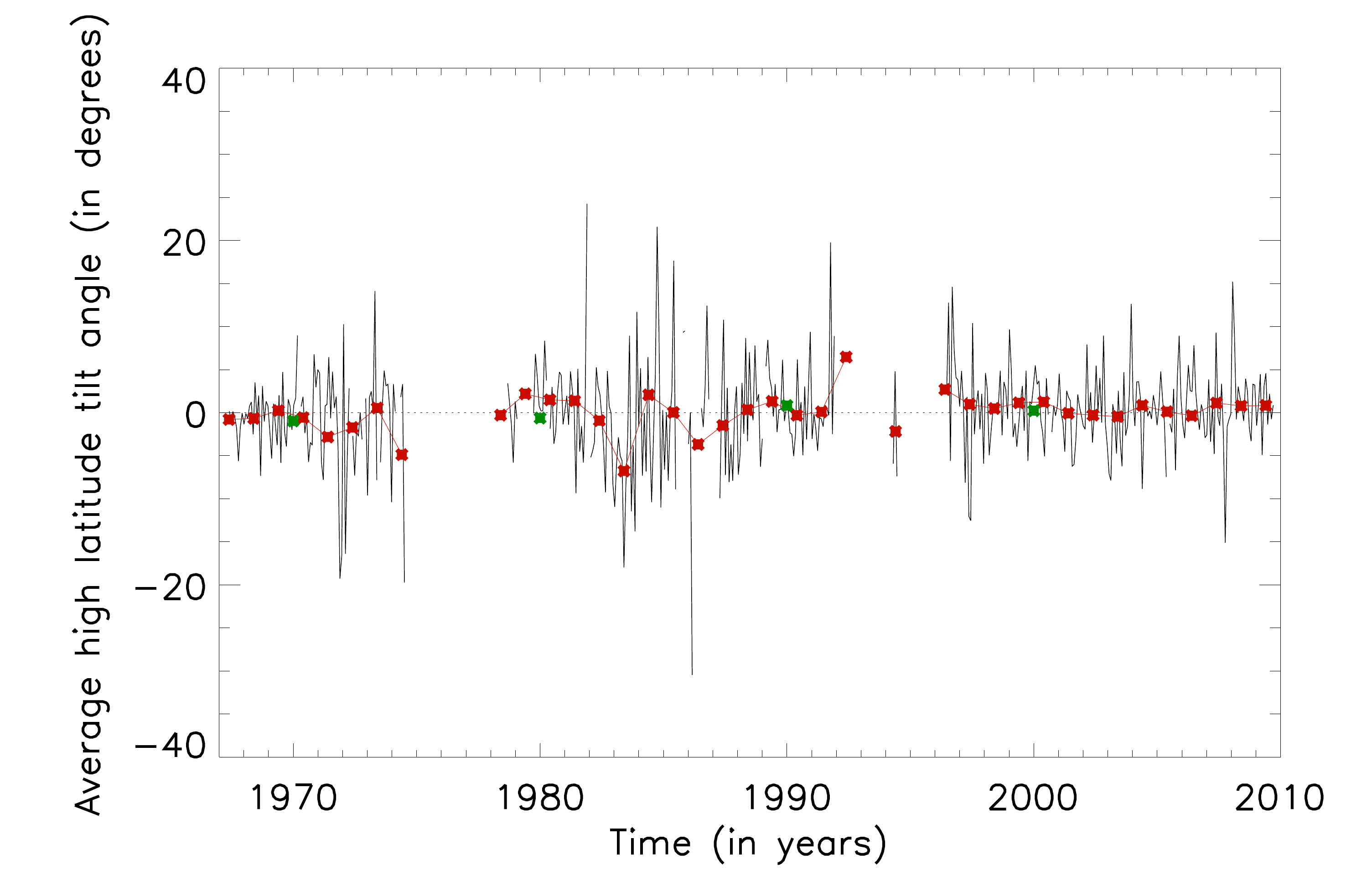}
\caption{Mean tilt angle variation of high latitude filament with time. Black line depicts the high latitude mean tilt angle variation with time. The red line shows the temporal variation of high latitude mean tilt angle averaged over 13 Carrington rotations. The green cross represents solar cycle average mean tilt angle.}
\label{tilt_avg_h}
\end{figure}

\begin{figure}[htb!]
\centering
\includegraphics[scale=1.0,angle=90]{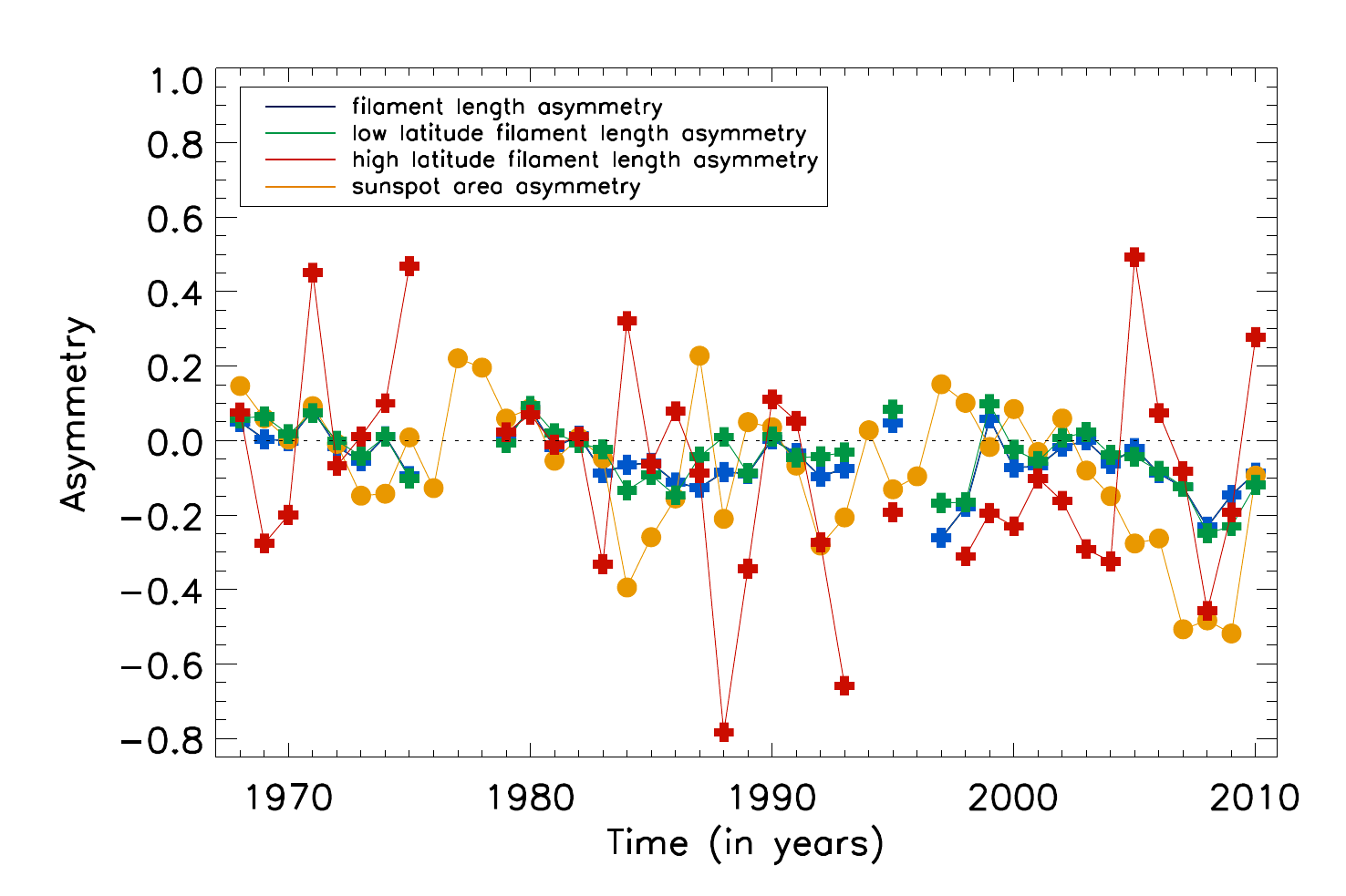}
\caption{Hemispheric asymmetry of filament and sunspot area.  The blue cross represents the north-south asymmetry of all filaments. The green cross represents the north-south asymmetry of low latitude filament. The red cross represents the north-south asymmetry of high latitude filament. The north-south asymmetry of sunspot area is depicted by filled orange circles.}
\label{ns_asymm}
\end{figure}

  \section{Data and methods}
 Patrick  S McIntosh, a scientist of NOAA’s Space Environment Center in Boulder, had drawn the Carrington maps of the Sun using the various ground as well as space-based satellite data from April 1967 till July 2009. The hand-drawn maps are archived and digitized by McIntosh project (a Boston College/NOAA/NCAR collaboration, funded by the NSF) at NOAA/NCEI. The data are stored in both image and fits format and also have been made available online\footnote{\url{https://www2.hao.ucar.edu/mcIntosh-archive/four-cycles-solar-synoptic-maps}}. For our analysis, we have used level 3 fits file from McIntosh archive \citep{2017IAUS..328...93G,SWE:SWE20537}.

  Level 3 image of the Carrington rotation number 1916 is presented in Figure \ref{level3}. The Carrington maps contain spatial information of various solar features namely, sunspots, plages, filaments, coronal holes and Polarity Inversion Lines (PILs). In this work, we have studied different properties of filaments. There are three big gaps in the data, first from June 1974 to July 1978, second from October 1991 to January 1994, and the final from April 1994 to May 1996.
We have reduced the level 3 fits files using Interactive Data Language (IDL) and obtained the spatial information of various solar features, namely, sunspots, plages, filaments, coronal holes, and Polarity Inversion Lines (PILs). In this study, we are interested in filaments alone. So we produce Carrington map that contains only the filament. In total available 442 Carrington maps during April 1967 to July 2009, we have detected 67373 filaments. We fit a straight line to each filament. The angle that the straight line makes with the equator is defined as the tilt angle of that particular filament. One sample image of Carrington map 1552 containing the only filament with the fitted straight line is shown in Figure \ref{sample_fil_fit}. The filaments are represented by green lines, and the fitted straight lines are represented by red lines. We observe that the fitting is good except for a few cases which can be ignored for statistical analysis. We calculate filament length using the following formula

\begin{equation}
 L = \sum_{n} \sqrt{R_{\odot}^{2}\delta\theta^{2}+R_{\odot}^{2}Cos^2 \theta \delta\phi^{2}}
\end{equation}

\noindent where L is filament's length, $R_{\odot}$ is the solar radius, the symbols $\theta$ and $\phi$ represent the latitude and longitude of a particular pixel respectively, and $n$ is the total number of pixels associated with the filament's structure. The quantities $\delta \theta$ and $\delta \phi$ are latitudinal and longitudinal differences between two adjacent pixels, respectively.
We add the lengths of all the filaments in a Carrington map to get the total length of the all filaments in it. 
The sunspot data are taken from the Royal Greenwich Observatory
(RGO) and US Air Force (USAF) - Solar Optical Observing Network (SOON) database.

\section{Analysis and Results}
Filament always appears in the vicinity of PILs \citep{1998SoPh..182..107M}. 
Filament formation is closely related to generation of magnetic field in the Sun.
Figure \ref{fil_no} represents the variation of filament number with time. The black line shows the variation of the total filament number with time. The green line depicts the variation of low latitude filament number with time. The red line represents the variation of high latitude filament number with time. The variation of sunspot area is depicted by the orange dashed line.
We notice a cyclic variation of the total filament number and low latitude filament number which is in phase with the sunspot area variation. The number of high latitude filaments is less, but they also show a cyclic variation \citep{2007MNRAS.376L..39L}.\\

The tilt angle of the active region (AR) is a very crucial parameter in the solar dynamo. The tilt angle of active region (AR) magnetic bipoles governs the conversion efficiency of the toroidal magnetic field to poloidal magnetic field \citep{2013MNRAS.432.2975T}. Filaments always form over PILs. Thus, the analysis of the tilt angle of filaments can give us important insights into the magnetic field generation mechanism inside the Sun. Here, we carry out a detailed analysis of the filaments' tilt angle. Figure \ref{hist_tilt} represents the histogram of the filament tilt angle. Out of a total 67373 filaments, 33415 filaments have positive tilt angle, and 33622 filaments have negative tilt angle.
So, in our database, there are 207 more filaments which have a negative tilt angle, when compared to filaments which have positive tilt angle.
Figure \ref{hist_pv} shows the latitudinal distribution of positive tilt angle filaments. Out of total 
33415 positive tilt angle filaments, 12113 filaments are in the northern hemisphere, and 21297 filaments are in the southern hemisphere. Figure \ref{hist_nv} shows the latitudinal distribution of negative tilt angle. Out of 33622 total negative tilt angle filaments, 20305 filaments are in the northern hemisphere, and 13315 filaments are in the southern hemisphere. So there exists a dominance of negative tilt angle in the northern hemisphere while positive tilt angle dominates in the southern hemisphere. The findings are consistent with earlier findings by \cite{2018ApJ...868...52M}. \\ 

Figure \ref{tilt_avg} depicts the variation of the mean tilt angle with time. We calculate mean of all tilt angles present in a particular Carrington map and get the mean tilt angle. The black curve in Figure \ref{tilt_avg} represents the variation of the mean tilt angle with time. The red line depicts the temporal variation of the mean tilt angle averaged over 13 Carrington rotations. We further calculate the average tilt angle in each solar cycle. The green cross represents solar cycle average mean tilt angle. We further investigate the lower latitude and higher latitude filaments' mean tilt angle separately. We define lower 
latitude filaments to be filaments situated within latitude $\pm 40^{\circ}$ ( $ |\theta |< 40^{\circ}$) and we define higher latitude filament as filaments having latitude either greater than 
$50^{\circ}$ or filaments having latitude less than $- 50^{\circ}$ ($ |\theta |> 50^{\circ}$). The total number of low latitude filaments is 51210, and the total number of high latitude filaments is 7564. Figure \ref{tilt_avg_lw} depicts the variation of low latitude mean tilt angle with time. The black line shows the low latitude mean tilt angle variation with time. The red line shows the temporal variation of low latitude tilt angle averaged over 13 Carrington rotations. The green cross represents solar cycle average mean tilt angle in low latitude. Figure \ref{tilt_avg_h} depicts the variation of high latitude mean tilt angle with time. The black line shows high latitude mean tilt angle variation with time. The red line shows the temporal variation of high latitude tilt angle averaged over 13 Carrington rotations. The green cross represents solar cycle average mean tilt angle in high latitude.

Filament formation is not symmetric in two hemispheres of the Sun. Earlier studies reported the north-south asymmetry of filament number \citep{2015ApJS..221...33H,2015RAA....15...77K,2010NewA...15..346L,2003NewA....8..655L}, but we believe that total filament length captures the magnetic field generation information in the Sun better than the filament number \citep{2018ApJ...868...52M}. So we use the total filament length in each Carrington rotation to calculate the north-south asymmetry of the filament. We define north-south asymmetry of sunspot area $A_{sp}$ as
\begin{equation}
A_{sp}=\frac{N_{sp}-S_{sp}}{N_{sp}+S_{sp}}
\end{equation}
 \noindent where $N_{sp}$ and $S_{sp}$ are the total area of the sunspots in the northern and the southern hemispheres, respectively. If $A_{sp} > 0$, the total sunspot area in the northern hemisphere dominates than that in the southern hemisphere and if $A_{sp} < 0$, the total sunspot area in the southern hemisphere dominates than that in the northern hemisphere.
We define north-south asymmetry of the filament ($A_{fil}$) as
\begin{equation}
A_{fil}=\frac{N_{fil}-S_{fil}}{N_{fil}+S_{fil}}
\end{equation}
 \noindent where $N_{fil}$ and $S_{fil}$ are the total filament length in northern and southern hemisphere, respectively. If $A_{fil} > 0$, the total filament length in the northern hemisphere dominates than that in the southern hemisphere  and if $A_{fil} < 0$, the total filament length in the southern hemisphere dominates than that in the northern hemisphere.

 Figure \ref{ns_asymm} depicts north-south asymmetry of the filament length and sunspot area for different latitude criteria. The blue cross represents the north-south asymmetry of all filaments. The green cross represents the north-south asymmetry of low latitude filaments. The red cross represents the north-south asymmetry of high latitude filaments.
The north-south asymmetry of sunspot area is shown by filled orange circles.
The temporal variation of north-south asymmetry of the total filaments and low latitude filaments show a sunspot area north-south asymmetry like behaviour. The north-south asymmetry of high latitude filament fluctuates and does not show any correspondence to sunspot area north-south asymmetry. However, the north-south asymmetry of high latitude filaments shows dominance of northern hemisphere in solar cycle 20 whereas, in the solar cycles 21, 22 and 23, the southern hemisphere shows a dominant character (Li et al. 2009).

\section{Summary and Conclusion}
We have analyzed 442 Carrington maps from MacIntosh database and detected 67373 filaments. Each filament is fitted with a straight line, and the tilt angle (angle which the straight line makes with equator) is estimated. We have noticed a cyclic variation of the total filament number as well as in low latitude and high latitude filament numbers (see Figure \ref{fil_no}). A detailed analysis of tilt angles of the filament is carried out in this work. Figure \ref{sample_fil_fit} shows the example of fitting of the filament (depicted by green lines) with the straight line (depicted by red straight lines) for the Carrington map 1552. The histogram of the tilt angle of all filaments are plotted (see Figure \ref{hist_tilt}), and we have found an excess of 207 negative tilt angle filaments as compared to the positive tilt angle filaments. Histogram of the latitude of positive tilt angle filament reveals that positive tilt angle is dominant in the southern hemisphere (see Figure \ref{hist_pv}). Histogram of the latitude of negative tilt angle filament reveals that negative tilt angle is dominant in the northern hemisphere (see Figure \ref{hist_nv}). The findings are consistent with an earlier report \citep{2018ApJ...868...52M} and contradict our expectation of positive tilt angle dominance in the northern hemisphere and negative tilt angle dominance in the southern hemisphere according to Hale's polarity law in the two hemispheres \citep{1919ApJ....49..153H}. It can happen due to the contribution of inter-AR filament formation geometry explained by \citep{2001ApJ...558..872M}. Another possible explanation is that more filaments are formed from large-scale magnetic field structure than from the intra-AR filament \citep{2018ApJ...868...52M}.  We examined the variation of mean tilt angle (mean of the tilt angle of all filaments in a Carrington map) and cycle-averaged tilt angle (see Figure \ref{tilt_avg}). We also studied mean tilt angle variation in low latitude  ($ |\theta |< 40^{\circ}$)  and high latitude ($ |\theta |> 50^{\circ}$) (see Figure \ref{tilt_avg_lw} and Figure \ref{tilt_avg_h}). 
Similar analysis carried out by \cite{2016SoPh..291.1115T} (from different database) had more extended and continuous data. So we are skeptical about comparing our result to their findings.
The north-south asymmetry of the total filament, low latitude filament, and high latitude filament reveals a sunspot area north-south asymmetry like behaviour in case of the total filament and low latitude filament (see Figure \ref{ns_asymm}). The north-south asymmetry of high latitude filament fluctuates and does not show any correspondence to sunspot area north-south asymmetry. However, the north-south asymmetry of high latitude filaments shows a dominance of northern hemisphere in solar cycle 20 whereas in the solar cycles 21, 22 and 23 the southern hemisphere shows a dominant character \citep{2009MNRAS.394..231L}.

\section*{Acknowledgement}
We are thankful to Patrick S. McIntosh, from NOAA's Space Environment Center in Boulder
for developing the archive of hand-drawn Carrington maps using both ground-based and satellite observations. We are also grateful to the team of McIntosh
archive project (a Boston College/NOAA/NCAR collaboration, funded by the NSF), NOAA National Centers for Environmental Information for creating a digital archive of McIntosh Carrington maps and make it available online. R.M. acknowledges the grant from the University Grant Commission, Government of India and CESSI, IISER Kolkata, India.\\

\bibliographystyle{raa}
\bibliography{bibtex}

\end{document}